\documentclass[a4paper,12pt,fleqn]{article}
\usepackage[official]{eurosym}  %Package fr Euro Zeichen
\usepackage[latin1]{inputenc}  \usepackage[T1]{fontenc}
\usepackage{lscape}
\usepackage{tabularx}     
\usepackage{pstricks}    
\usepackage{rotating}
\usepackage{amsfonts,amsmath,amsthm,amssymb,dsfont}
\usepackage{marginnote}
\usepackage{rotate}
\usepackage{array,multirow}
\usepackage{footnote}
\usepackage{url}
\makesavenoteenv{tabular}
\usepackage[arrow,matrix,curve]{xy} 	
\title{The Influence of Seed Selection on the Solvency II Ratio}
\author{Quinn Culver \and Dennis Heitmann \and Christian Wei\ss}
\date{\today}

\newtheorem{thm}{Theorem}
\newtheorem{defi}{Definition}

\makeindex

\begin{document} 

\maketitle

\begin{abstract} This article contains the first published example of
  a real economic balance sheet where the Solvency II ratio
  substantially depends on the seed selected for the random number
  generator (RNG) used.  The theoretical background and the main
  quality criteria for RNGs are explained in detail.  To serve as a
  gauge for RNGs, a definition of true randomness is given. Quality
  tests that RNGs should pass in order to generate stable results when
  used in risk management under Solvency II are described.
\end{abstract}

%\tableofcontents

\paragraph{Introduction.} Most German insurance companies use the
standard formula for risk capital calculation, implying in almost all
cases the use of the simulation model
\textit{Branchensimulationsmodell} (BSM), provided by the German
Insurance Association (GDV) in order to evaluate the Best Estimate
Liabilities (BEL), Own Funds (OF) and the Solvency Capital Requirements (SCR) within the Solvency II framework. Some insurance companies, especially the European-wide capital stock companies, use an internal model, %i.e. \textit{Replication Portfolio}\footnote{A replicating portfolio is a pool of assets designed to reproduce (replicate) the cash flows or market values of a pool of liabilities across a large number of stochastic scenarios.} or \textit{LSMC}\footnote{Least Squares Monte Carlo}. 
In this paper we consider exlusively the BSM. According to legislation (see \cite{EIO15}), the methodology for calculating the BEL has to be accurate and robust. In particular for life insurance a very important component of the calculation of BEL is the Economic Scenario Generator (ESG). Here, Guidelines 55-59, \cite{EIO15}, explicitly define requirements for its quality.%Among the most important
%of these choices is the Economic Scenario Generator (ESG). 
The reliability of a Monte Carlo simulation of the BSM (or any
other internal model) does not only depend on the quality of the model
from a classical actuarial point of view but also on how well the
underlying ESG performs. In this article, we put our focus on a
mathematically challenging aspect within the ESG, namely the random
numbers used therein serving as driver of stochasticity. Guideline 59, \cite{EIO15}, explicitly demands the proper testing of random number generators used in the ESG.  The ESG
provided by the GDV is implemented in Excel and is based on a combined
linear congruential generator (LCG) of Wichmann and Hill (see
\cite{Wie82}) in order to generate the underlying random
numbers. % Therefore, it is worth analyzing the quality of the random number generator in Excel. Therefore, let us consider an example. %Solvency II requires an economic balance sheet representing assets, own funds and liabilities at a given date. Due to the fact that the TVOG is defined by the difference of the stochastic present value of future profits (PVFP) and the certainty equivalent (CE) PVFP, it is hence crucial to generate suitable market consistent scenarios. This makes it crucial to generate suitable market consistent scenarios. According to the Delegated Acts \cite{EC15}, Article 229, the output of the model is required to be "stable in relation to changes in the input data that do not correspond to a relevant change of the risk profile of the insurance or reinsurance undertaking".
\paragraph{Example.} To demonstrate the impact of the random number generator, let us consider an example of a typical German life insurance company using the BSM with a strong focus on endowment business in their portfolio.  From a regulatory point of view, it is required to have stable results depending only minimal on the random number generator of the ESG. However, our data shows a significant impact of one of these settings within the ESG, namely the seed selection, on relevant financial statement data in the Economic Balance Sheet and ultimately the
Solvency II ratio. The endowment business is typical for the German market and the BSM was mainly developed for its valuation. %In addition, we have considered the impact of the transitional provisions\footnote{The transitional provisions can be
  %applied in Germany from 2016 until 2032 with a linear decreasing
  %adjustments of the BEL in order to facilitate the transition from
  %the Solvency I regime to the Solvency II regime.} (R\"uckstellungstransitional, RT) which is widely used in the German market. 
  Based on our data the Solvency II ratio for YE 2016 was in the range of typical life insurance companies in the German market.  Since focus on the stability of the results, we look at the relative changes and differences in percentage points and not absolute figures. In Table~1 a short comparison of the relevant financial data of two seeds is given.%\footnote{The delta of  expected guarantees is 0.6\% and not listed here due to the fact  that the absolute values are not significant.} 
  These two seeds are two of in total 30 seeds analysed, and show typical differences and not the maximal as observed in the set.  %Three types of relative changes in the Solvency
\begin{center}
\begin{tabular}{|l|l|r|}
	\hline 
	%\small{Relevant financial information} & Delta Options $[\%]$ & \ \ \ 11.6\\ \cline{2-3}
	%& Delta TVOG $[\%]$ & -10.1\\ \cline{2-3}
	\small{Relevant financial information} & Delta Own Funds  $[\%]$ & -8.2\\ \cline{2-3}
	& Delta SCR  $[\%]$ & 7.2\\ \cline{2-3}
	& Delta Risk Margin $[\%]$ & 9.9\\ \hline \hline
	%\small{BSM} & Delta SII Ratio $[\%]$ & -14.4\\ \cline{2-3}
	%& Delta SII Ratio [pp] & -9.9\\ \hline \hline
	%\small{Risk reduction due to deferred tax} & Delta SII Ratio $[\%]$ & -14.4\\ \cline{2-3}
	%& Delta SII Ratio [pp] & -17.0\\ \hline \hline
	\small{Final Solveny II ratios} & Delta SII Ratio $[\%]$ & -14.2\\ \cline{2-3}
	& Delta SII Ratio [pp] & -35.0\\ \hline
 	%\small{and transitional provisions} & Delta SII Ratio [pp] & -35.0\\ \hline
\end{tabular}\\[12pt]
Table~1. Comparison of relevant financial data among two seeds.
\end{center}

As we can assume that the BSM model works properly, there are essentially two explanations why the results in the example differ by such a large amount: Either the rates of convergence of the Monte Carlo simulation distinguish or one of the seeds produces a \textit{more random} sequence than the other. %While there are many standard techniques available to increase the rate of convergence by variance reduction \textit{after} the random sequence has been generated (see e.g. \cite{Gla03}), the assessment of randomness needs to be based purely on the  sequence under consideration.

\paragraph{What does randomness mean?} % When we want to talk about
% assessing \textit{randomness}, we have to at first define in a
% mathematical precise manner what we understand by it.

% We will restrict discussing randomness in the unit interval $[0,1]$
% with the uniform distribution. This is really not a restriction.

%In probability theory, the term ``random'' is used to describe a
%property that holds almost-surely (that is, with probability one).
The term ``random'' describes the absence of patterns and predictability. For
example, it would be correct to say that if $x$ is chosen \emph{at
  random} according to the uniform distribution from the unit interval
$[0,1]$, then (by the strong law of large numbers), it satisfies the
following property:
\begin{itemize}
  \item[P1.] Each of the digits $0$-$9$ shows up in $x$'s decimal
  expansion with frequency $\frac{1}{10}$; e.g.\
  \begin{displaymath}
    \lim_{n\to \infty} \frac{\text{the number of 7's in the first $n$ digits of
        $x$}}{n} = \frac{1}{10}.
  \end{displaymath}
\end{itemize}
P1 is one example of a randomness property but its satisfaction is
certainly not sufficient to guarantee randomness; there are numbers,
like $0.\overline{0123456789}$, that are intuitively nonrandom, but
satisfy P1.

So instead of satisfying only P1, one might require it also satisfy
some other (almost-sure) property, P2. But then an example of something intuitively nonrandom
satisfying P1 and P2 can be exhibited.

It is then natural to attempt to define randomness by considering only
$x$'s that satisfy \emph{all} randomness properties. The problem with
this approach is that there are too many randomness properties, since
the property ``not equal to $x$'' holds almost-surely and hence is a
randomness property. If a random number were required to satisfy that
property for each $x$, there would then be nothing left to call
random.

The theory of computation gives a natural way to restrict the kind of
randomness property; only those properties that can be effectively
checked by a computer. (For sufficiently complicated $x$'s, a computer
cannot check ``not equal to $x$''.)

Definition~\ref{def:MLrandomness} abstracts the notion of a randomness
property by exploiting \emph{outer regularity}; any probability-zero
event can be covered by open set of arbitrarily small probability. Any
almost-sure property that one would want a truly random $x\in [0,1]$
to satisfy is representable as a so-called Martin-L\"{o}f
test. %INCOMPLETE: CITE? 

\begin{defi}[\cite{Nie09}] \label{Def1}
  A \emph{(Martin-L\"{o}f) test} for randomness is a sequence
  $U_{1}, U_{2},\ldots$ of open subsets of $[0,1]$, with the following
  properties
  \begin{itemize}
    \item $U_{n} = \bigcup_{k} (a_{n,k},b_{n,k})$ for rational numbers
    $a_{n,k}, b_{n,k}$,
    \item $\Pr(U_{n})\leq 2^{-n}$,
    \item there is an (abstract) computer program that, given inputs
    $n$ and $k$, outputs $a_{n,k}$ and $b_{n,k}$.
  \end{itemize}
  A number $x\in [0,1]$ \emph{passes the test} if
  $x \notin \bigcap U_{n}$ and $x$ is called \emph{(Martin-L\"{o}f)
    random} if it passes every such test.
  \label{def:MLrandomness}
\end{defi}

Furthermore, Definition~\ref{def:MLrandomness} gives a definition that is
inherently not practical; finitely many digits of a given $x\in [0,1]$
do not determine its randomness. However, from a practioner's perspective only finite samples are of relevance.  As John Von Neumann said, ``Any one who considers arithmetical methods of producing random digits is, of
course, in a state of sin.'' (See \cite{wiki:JVNquote})
% INCOMPLETE: CITE? add footnote about arithmetical=computational

\paragraph{Random Number Generators.} Before we can explain how to test practically, if a sequence at hand is random or not we have to better understand how a computer produces random numbers and what is meant by \textit{seed selection}. At the core of every Monte Carlo simulation there is a sequence of random variables $X_1, X_2,...$ which is
\begin{itemize}
 \item[(i)] uniformly distributed in the interval $[0,1]$ and 
 \item[(ii)] satisfies the property that the $X_i$ are mutually independent.\footnote{Property (i) is not a restriction since there are many ways to transform random variables uniformly distributed on $[0,1]$ to arbitrary distributed random variables, compare \cite{Gla03}, Chapter~2.2. These methods are applied by the economic scenario generators which are used by insurance companies.} 
\end{itemize} 
A random number generator is a mechanism producing such a
sequence. The economic scenario generator is fed by these numbers. One
of the most ubiquitous random number generators is \textbf{Mersenne-Twister}. Like most other commonly used random number generators it is based on congruence calculations.\footnote{An example of a random number generator not relying on congruence is John von Neumann's middle-square method which is for several reasons not a good method in practice.} Since we only intend to familiarize the reader with the underlying ideas but not to treat the topic in its full generality (see e.g. \cite{Gla03}, Chapter~2, \cite{Nie92}, Chapter~7 and most importantly \cite{PTVF07}, Chapter~7 for more details), we explain here only the simple linear congruential generator introduced by Lehmer in \cite{Leh51} which suffices for an overall understanding.\\[12pt]
At first we choose a large integer $m$, called the \textbf{modulus},
an integer $a$, called the \textbf{multiplier}, with $0 < a < m$, a
number $c$, called the \textbf{increment}, with $\gcd(c,m) = 1$ and a
\textbf{starting value} $Y_0$ with $0\leq Y_0 < m$. A sequence of
numbers is obtained by the recurrence
$$Y_{n+1} = (aY_n + c) \mod m, \qquad n \geq 0.$$
From that a sequence in $[0,1]$ can be derived by taking
$X_{n+1} = \tfrac{Y_{n+1}}{m}$. In fact, a sophisticated choice of the
numbers $a,c,m,Y_0$ is essential for getting a good random
sequence. This process is called \textbf{seed selection}.\footnote{For
  certain seeds, like $m = 10$, $Y_0=a=c=7$, the resulting sequence
  $X_{n}$ is obviously not random, see \cite{Knu98}.} Note that the
sequence $Y_N$ will repeat after applying the recurrence a certain
number of times. The repeating cycle is called \textbf{period}. Of
course, it is a desirable property that the period is as great as
possible. There is a well-known theorem stating that the period is at
most $m$ and it can also be described precisely when this is the
case.
\begin{thm} \textbf{(see \cite{Knu98}, Chapter 1, Theorem A)} The
  linear congruential sequence defined by $m, a, c$ and $Y_0$ has
  period length $m$ if and only if
	\begin{itemize}
		\item[(i)] $c$ is relatively prime to $m$,
		\item[(ii)] $b = a-1$ is a multiple of $p$ for every prime $p$ dividing $m$,
		\item[(iii)] $b$ is a multiple of $4$ if $m$ is a multiple of $4$.
	\end{itemize}	
\end{thm}
 
\paragraph{Quality Criteria for Random Number Generators.} Since the
random number generators commonly used in industry are based on
deterministic algorithms\footnote{Indeed, it is necessary that the
  algortihms be deterministic because of demands by the Solvency II delegated acts \cite{EC15}.} which are similar to the linear congruential method they inherit periodicity and do not produce independent random variables in the mathematical sense. Therefore, it is more appropriate to speak of \textbf{pseudorandom numbers} in this context. The main questions of the remainder of our article are now if and how we can distinguish pseudorandom numbers and random numbers. In other words: given one sequence of pseudorandom numbers and one sequence of random number can we decide statistically which one was made up by a computer? The less possible it is to detect that a sequence of numbers has been created by a computer the better the random number generator (or the random seed) is deemed.\\[12pt]
In the following we assume that we have a finite sequence of numbers
$x_1,\ldots,x_N$ in $[0,1]$ at hand. The present article does not aim
to give an exhaustive list of all known tests but should rather be
regarded as a selection of meaningful and functional tests the authors consider to be
particularly intuitive or easy-to-implement in practice. For a more
comprehensive overview we refer to \cite{ES07} and \cite{Knu98}. While
reading this paragraph the reader should always keep in mind
Definition~\ref{Def1} which makes clear that no single test or even
(finite) battery of tests can guarantee that a random number generator
is indeed a good mimic of randomness. Still these tests can improve
our confidence on the random number generator under consideration.
\subparagraph{Global Uniformity Test.} A very simple starting point is
to calculate the empirical mean and variance of the finite sequence
$x_1, \ldots,x_N$ and compare it to the values $1/2$ and $1/12$
expected theoretically. This can be done in a statistical precise way
by applying Student's t-test and Levene's test. Moreover, the
empirical distribution may in a next step be compared graphically to
uniform distribution. Finally a goodness-of-fit test like
Kolmogorov-Smirnov test, Chi-squared test or Anderson-Darling test may
be used to judge statistically if the empirically observed and
theoretically desired distribution coincide globally. Note that only a
very bad random number generator would fail these basic tests.
\subparagraph{Permutation Test.} We choose an arbitrary integral
number $2 \leq k \ll N-1$ and consider the tuples
$(x_n,x_{n+1},\ldots,x_{n+k-1})$ for $n=1,\ldots,N-k$. All the $k!$ relative orderings\footnote{For instance, the relative ordering of the $4$-tuple $(0.8,0.1,0.2,0.05)$ ist $(4,2,3,1)$.} among the entries of a generic $k$-tuple should be equiprobable, i.e. have probability $\tfrac{1}{k!}$. By counting the empirical frequency of the orderings in the sample and applying a goodness-of-fit test we may judge if the null-hypothesis that the relative orderings are equidistributed has to be rejected or not.  

\subparagraph{Serial Test.} A uniformly distributed sequence on $[0,1]$ is equally dense everywhere. Even $l$-tuples of successive numbers should be uniformly and independently distributed on $[0,1]^l$. The serial test analyses if this property is satisfied. For that purpose we at first partition each copy of $[0,1]$ into $d$ equally long pieces for some integer $d \geq 2$. By this construction we obtain $k := d^l$ subcubes of
volume $\tfrac{1}{k}$ and each subcube is therefore expected to contain $\lambda:=\tfrac{N}{k}$ elements of the finite sequence. The
serial test measures the discrepancy between the empiric counts $Y_j$ of the subcubes and $\lambda$. The corresponding test statistics is
given by 
$$X^2 := \sum_{j=0}^{r-1} \frac{(Y_j-\lambda)^2}{\lambda}$$
and it is approximately Chi-square distributed with $k-1$ degrees of freedom. Hence, we may at the very end apply a Chi-square test to see if equidistribution holds statistically. The crucial point about the serial test is the choice of $k$. In order to have a (good) approximate Chi-square distribution it is often recommended to have $N/k \geq 5$. This gives an upper bound on $d$ because otherwise the number of categories would be too large and the test would not be exact enough (see \cite{Knu98}, 3.3.2).
\subparagraph{Birthday Spacings Test.} The birthday spacings test is a refinement of the serial test. Its name is inspired by the birthday ``paradox'' which states that the likelihood that some pair of 23 randomly chosen people has the same birthday is $50 \%$.\footnote{In the language of the birthday paradox the length of the sub-sequence $n$ corresponds to the number of birthdays and the number of cells $k$ corresponds to the number of days in a year.} As for the serial test we decompose $[0,1]$ into $k$ cells and consider $n$ points of the sequence. Let $I_1 \leq I_2 \leq \ldots \leq I_n$ be the cell numbers where these $n$ points fall and let $Y$ be the number of collissions. It is well-known that $Y$ is approximately Poisson distributed with mean $n^3 / (4k)$ (see \cite{ES07}). A Chi-squared test can then be applied to the $N/n$ sample values of $Y$.
\begin{center}
\includegraphics[scale=0.4]{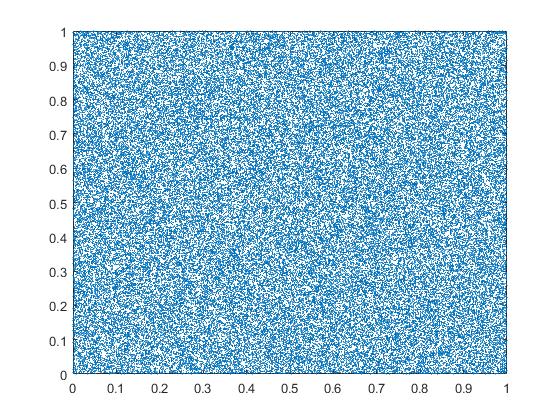}\\
Figure~1. Pairs $(x_k,x_{k+1})$ of pseudorandom numbers.
\end{center}
\subparagraph{Spectral Test.} While all tests so far were driven by statistics we complete our list by a purely geometrical test. Figure~1 and Figure~2 were generated using a pseudorandom number sequence similar to the one implemented in Excel.\footnote{In fact, we used the linear congruential method with $m = 2^{18}, a = 4649$ and $c=819$. This example has been chosen because the described effects become visible already in lower dimensions.}  They show the sets consisting of all pairs $(x_k,x_{k+1})$ respectively triples $(x_k,x_{k+1},x_{k+2})$. While Figure~1 does not seem to be particularly remarkable at first sight, Figure~2 is astonishing since all points lie on a few planes. 
\begin{center}
	\includegraphics[scale=0.4]{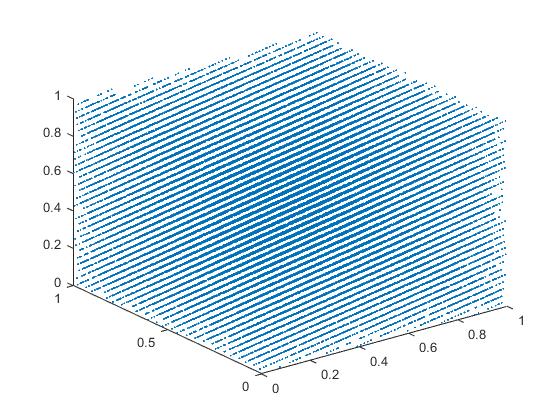}\\
	Figure~2. Triples $(x_k,x_{k+1},x_{k+2})$ of pseudorandom numbers.
\end{center}
On the other hand random number generators are neither expected nor desired to produce any symmetric geometric structures. This is a drawback of all random number generators which rely on linear congruence as we will explain now. Let $1 / \nu_2$ be the maximum distance between lines, taken over all families of parallel straight lines that cover all points $(x_k,x_{k+1})$. The greater $1 / \nu_2$ is, the fewer lines suffice to cover all points. In other words, the greater $\nu_2$ is the less the random number generator produces symmetric structures. Hence, $\nu_2$ is the two-dimensional \textbf{accuracy} of the random number generator. The concept of accuracy $\nu_d$ can easily be carried over to higher dimensions $d$, e.g. in dimension $3$ we replace lines by planes. Another interpretation of the numbers $\log \nu_d$ is that they measure how many digits of $d$-tuples may be regarded as being independent. The difference between pseudorandom numbers and random numbers truncated to multiples of some $1 / \nu$ is that random number sequences have approximately the same accuracy in all dimensions. It has been suggested in the literature to accept a random number generator for application whenever $\nu_d \geq 2^{30/d}$ for $2 \leq d \leq 6$. Practical implementation of spectral test demands some lines of programing code but can still be achieved with reasonable effort (see \cite{Knu98} for details). The number theory behind the spectral test is explained wonderfully in \cite{Kon03}.  %For pseudorandom numbers with period length $m$ there are only $m$ points in the $d$-dimensional cube such that the $d$-dimensional accuracy is bounded by $m^{1/d}$.\footnote{To be more precise, this is only true if the points form a regular grid, see \cite{Knu98}.} The numbers $\log \nu_d$ are a measure for how many digits of $d$-tuples may be regarded as being independent. 
 %However, we see that the pseudorandom numbers have certain dependencies. 
\paragraph{Conclusion.} This paper has revealed a significant impact of seed selection on the Solvency II ratio. Moreover, we explained how randomness can be defined in a mathematical rigorous sense and showed how statistical and geometrical tests can be used to distinguish good seeds of an random number generator from bad ones. We strongly suggest that these quality criteria for random number generators will be considered by insurance undertakings, regulators and external auditors as well in the future. Despite proving appropriateness of the interest rate model\footnote{The Hull-White model is used in the ESG of the BSM.}, the software used for calculating the provision for future policy benefits, the management rules and so forth, the random number generator should also be added in the framework for reasons of adequacy consideration.

\paragraph{Acknowledgments.} The authors wish to thank two anonymous actuarial employees of a German insurance company for their calculations and providing the data. Furthermore, we are grateful to referee for his useful comments.

\textsc{Quinn Culver, Fordham University, 407 John Mulcahy Hall Bronx, NY 0458-5165}\\
\textit{E-mail address:} \texttt{qculver@fordham.edu}

\textsc{Dennis Heitmann, BSP Business School Berlin - Campus Hamburg, Am Kaiserkai 1, D-20457 Hamburg}\\
\textit{E-mail address:} \texttt{dennis.heitmann@bsp-campus-hamburg.de}

\textsc{Christian Wei\ss, Hochschule Ruhr West, Duisburger Str. 100, D-45479 M\"ulheim an der Ruhr}\\
\textit{E-mail address:} \texttt{christian.weiss@hs-ruhrwest.de}


\begin{thebibliography}{xxx}	
	\bibitem[EC15]{EC15} European Commission, ``Commission Delegated Regulation (EU) 2015/35'' (2015).
	\bibitem[EIO15]{EIO15} EIOPA,  ``Guidelines on the valuation of technical provisions'' (2015).
	\bibitem[ES07]{ES07} L'Ecuyer, P., Simard, R.: ``TestU01: A C Library for Empirical Testing of Random Number Generators'', ACM Transactions on Mathematical Software, 33 (4), Article 22 (2007).
	\bibitem[Gla03]{Gla03} Glasserman, P.: ``Monte Carlo Methods in Financial Engineering'', Springer (2003).
	\bibitem[Knu98]{Knu98} Knuth, D.E.: ``The Art of Computer Programming'', Vol 2: Seminumerical Algorithms, Addison-Wesley (1998).
	\bibitem[Kon03]{Kon03} Kontorovich, A.: ``From Apollonius to Zaremba: Local-Global Phenomena in Thin Orbit'', Bull. AMS 50,, No. 2, 187--228 (2013).
	\bibitem[Leh51]{Leh51} Lehmer D.H.: ``Mathematical methods in large-scaling computing units'', Proc. 2nd Sympos. of Large-Scale Digital Calculating Machinery, Harvard University Press, 141--146 (1951).
	\bibitem[Nie92]{Nie92} Niederreiter, H.: ``Random Number Generation and Quasi-Monte Carlo Methods'', Number 63 in CBMS-NSF Series in Applied Mathematics, SIAM, Philadelphia (1992).
        \bibitem[Nie09]{Nie09} Nies, Andr{\'e}: ``Computability and randomness'', Oxford University Press, 2009.
	\bibitem[PTVF07]{PTVF07} Press, W, Teukolsky, S., Vetterling, W., Flannery, B.: ``Numerical Recipes: The Art of Scientific Computing'', Cambridge University Press (2007).
	\bibitem[Wie82]{Wie82} Wichmann, B.A., Hill, I.D..: ``An
        efficient and portable pseudo-random number generator'',
        App. Statis. 31, 188-190 (1982).
    \bibitem[VNWiki]{wiki:JVNquote} Wikiquote: ``John von Neumann --- Wikiquote{,}'', 25 April 2017, Retrieved December 20, 2017 from
        \url{https://en.wikiquote.org/w/index.php?title=John_von_Neumann&oldid=2246778}.
\end{thebibliography}
\end{document}